\documentclass[aps,prl,preprint,floatfix,superscriptaddress]{revtex4}

\usepackage{graphicx}%
\usepackage{dcolumn}%

\newcommand{\ps}{e$^{+}$}
\newcommand{\el}{e$^{-}$}

\begin{document}
\title{Observation of an anomalous positron abundance in the
cosmic radiation}

\author{O. Adriani}
\affiliation{Physics Department of
University of Florence,  
 I-50019 Sesto Fiorentino, Florence, Italy}
\affiliation{INFN, Sezione di Florence,  
 I-50019 Sesto Fiorentino, Florence, Italy}
\author{G. C. Barbarino}
\affiliation{Physics Department of University of
Naples 
``Federico II'',  I-80126 Naples, Italy}
\affiliation{INFN, Sezione di Naples,  I-80126 Naples, Italy}
\author{G. A. Bazilevskaya}
\affiliation{Lebedev Physical Institute, Leninsky Prospekt 53, RU-119991
Moscow, Russia}
\author{R. Bellotti}
\affiliation{Physics Department of
University of Bari, I-70126 Bari, Italy}
\affiliation{INFN, Sezione di Bari, I-70126 Bari, Italy}
\author{M. Boezio}
\affiliation{INFN, Sezione di Trieste, I-34012
Trieste, Italy}
\author{E. A. Bogomolov}
\affiliation{Ioffe Physical Technical Institute,  RU-194021 St. 
Petersburg, Russia}
\author{L. Bonechi}
\affiliation{Physics Department of
University of Florence,  
 I-50019 Sesto Fiorentino, Florence, Italy}
\affiliation{INFN, Sezione di Florence,  
 I-50019 Sesto Fiorentino, Florence, Italy}
\author{M. Bongi}
\affiliation{INFN, Sezione di Florence,  
 I-50019 Sesto Fiorentino, Florence, Italy}
\author{V. Bonvicini}
\affiliation{INFN, Sezione di Trieste,  I-34012
Trieste, Italy}
\author{S. Bottai}
\affiliation{INFN, Sezione di Florence,  
 I-50019 Sesto Fiorentino, Florence, Italy}
\author{A. Bruno}
\affiliation{Physics Department of
University of Bari,  I-70126 Bari, Italy}
\affiliation{INFN, Sezione di Bari, I-70126 Bari, Italy}
\author{F. Cafagna}
\affiliation{INFN, Sezione di Bari, I-70126 Bari, Italy}
\author{D. Campana}
\affiliation{INFN, Sezione di Naples,  I-80126 Naples, Italy}
\author{P. Carlson}
\affiliation{Physics Department of the Royal Institute of Technology
(KTH), SE-10691 Stockholm,  
Sweden}
\author{M. Casolino}
\affiliation{INFN, Sezione di Rome ``Tor Vergata'', I-00133 Rome, Italy}
\author{G. Castellini}
\affiliation{ IFAC,  I-50019 Sesto Fiorentino,
Florence, Italy}
\author{M. P. De Pascale}
\affiliation{INFN, Sezione di Rome ``Tor Vergata'', I-00133 Rome, Italy}
\affiliation{Physics Department
of University of Rome ``Tor Vergata'',  I-00133 Rome, Italy}
\author{G. De Rosa}
\affiliation{INFN, Sezione di Naples,  I-80126 Naples, Italy}
\author{N. De Simone}
\affiliation{INFN, Sezione di Rome ``Tor Vergata'', I-00133 Rome, Italy}
\affiliation{Physics Department
of University of Rome ``Tor Vergata'',  I-00133 Rome, Italy}
\author{V. Di Felice}
\affiliation{INFN, Sezione di Rome ``Tor Vergata'', I-00133 Rome, Italy}
\affiliation{Physics Department
of University of Rome ``Tor Vergata'',  I-00133 Rome, Italy}

\author{A. M. Galper}
\affiliation{Moscow Engineering and Physics Institute,  RU-11540 Moscow, Russia} 
\author{L. Grishantseva}
\affiliation{Moscow Engineering and Physics Institute,  RU-11540 Moscow, Russia} 
\author{P. Hofverberg}
\affiliation{Physics Department of the Royal Institute of Technology
(KTH), SE-10691 Stockholm,  
Sweden}
\author{A. Leonov}
\affiliation{Moscow Engineering and Physics Institute,  RU-11540 Moscow, Russia} 
\author{S. V. Koldashov}
\affiliation{Moscow Engineering and Physics Institute,  RU-11540 Moscow, Russia} 
\author{S. Y. Krutkov}
\affiliation{Ioffe Physical Technical Institute,  RU-194021 St. 
Petersburg, Russia}
\author{A. N. Kvashnin}
\affiliation{Lebedev Physical Institute,  RU-119991
Moscow, Russia}
\author{V. Malvezzi}
\affiliation{INFN, Sezione di Rome ``Tor Vergata'', I-00133 Rome, Italy}
\author{L. Marcelli}
\affiliation{INFN, Sezione di Rome ``Tor Vergata'', I-00133 Rome, Italy}
\author{W. Menn}
\affiliation{Physics Department
of Universit\"{a}t Siegen, D-57068 Siegen, Germany}
\author{V. V. Mikhailov}
\affiliation{Moscow Engineering and Physics Institute,  RU-11540
Moscow, Russia}  
\author{E. Mocchiutti}
\affiliation{INFN, Sezione di Trieste,  I-34012
Trieste, Italy}
\author{S. Orsi}
\affiliation{Physics Department of the Royal Institute of Technology
(KTH), SE-10691 Stockholm,  
Sweden}
\author{G. Osteria}
\affiliation{INFN, Sezione di Naples,  I-80126 Naples, Italy}
\author{P. Papini}
\affiliation{INFN, Sezione di Florence,  
 I-50019 Sesto Fiorentino, Florence, Italy}
\author{M. Pearce}
\affiliation{Physics Department of the Royal Institute of Technology
(KTH), SE-10691 Stockholm,  
Sweden}
\author{P. Picozza}
\affiliation{INFN, Sezione di Rome ``Tor Vergata'', I-00133 Rome, Italy}
\affiliation{Physics Department
of University of Rome ``Tor Vergata'',  I-00133 Rome, Italy}
\author{M. Ricci}
\affiliation{INFN, Laboratori Nazionali di Frascati, Via Enrico Fermi 40,
I-00044 Frascati, Italy}
\author{S. B. Ricciarini}
\affiliation{INFN, Sezione di Florence, 
 I-50019 Sesto Fiorentino, Florence, Italy}
\author{M. Simon}
\affiliation{Physics Department
of Universit\"{a}t Siegen, D-57068 Siegen, Germany}
\author{R. Sparvoli}
\affiliation{INFN, Sezione di Rome ``Tor Vergata'', I-00133 Rome, Italy}
\affiliation{Physics Department
of University of Rome ``Tor Vergata'',  I-00133 Rome, Italy}
\author{P. Spillantini}
\affiliation{Physics Department of
University of Florence,  
 I-50019 Sesto Fiorentino, Florence, Italy}
\affiliation{INFN, Sezione di Florence,  
 I-50019 Sesto Fiorentino, Florence, Italy}
\author{Y. I. Stozhkov}
\affiliation{Lebedev Physical Institute,  RU-119991
Moscow, Russia}
\author{A. Vacchi}
\affiliation{INFN, Sezione di Trieste,  I-34012
Trieste, Italy}
\author{E. Vannuccini}
\affiliation{INFN, Sezione di Florence, 
 I-50019 Sesto Fiorentino, Florence, Italy}
\author{G. Vasilyev}
\affiliation{Ioffe Physical Technical Institute, RU-194021 St. 
Petersburg, Russia}
\author{S. A. Voronov}
\affiliation{Moscow Engineering and Physics Institute,  RU-11540
Moscow, Russia}  
\author{Y. T. Yurkin}
\affiliation{Moscow Engineering and Physics Institute,  RU-11540
Moscow, Russia}  
\author{G. Zampa}
\affiliation{INFN, Sezione di Trieste,  I-34012
Trieste, Italy}
\author{N. Zampa}
\affiliation{INFN, Sezione di Trieste,  I-34012
Trieste, Italy}
\author{V. G. Zverev}
\affiliation{Moscow Engineering and Physics Institute,  RU-11540
Moscow, Russia}  

\date{\today}

\begin{abstract}
Positrons are known to be produced in interactions between cosmic-ray nuclei
and interstellar matter (``secondary production''). Positrons may, however,
also be created by dark matter particle annihilations in
the galactic halo or in the magnetospheres of near-by pulsars. The nature of
dark matter
is one of the most prominent open questions in science today. An
observation of positrons from pulsars would open a new observation window on
these
sources. Here we present results from the PAMELA satellite experiment on the
positron
abundance in the cosmic radiation for the energy range 1.5 - 100 GeV. Our
high energy data
deviate significantly from predictions of secondary production models,
and may constitute the first indirect evidence of dark matter particle
annihilations, or the first observation of positron production from
near-by pulsars. 
We also present evidence that solar activity significantly affects the
abundance of positrons at low
energies.



\end{abstract}

\maketitle

Measurements of cosmic-ray positrons (\ps) and electrons (\el) address a
number of questions in contemporary astrophysics, such as the nature
and distribution of particle sources in our galaxy, and the subsequent propagation of
cosmic-rays through the galaxy and the solar magnetosphere. 
Positrons are believed to be mainly created in secondary production
processes resulting from  
the interaction of cosmic-ray nuclei with the interstellar gas. 
A ratio of positron and electron fluxes ($\phi$), the positron fraction, 
\mbox{$\phi$(e$^+$) / ($\phi$(e$^+$) + $\phi$(e$^-$))}, can be used to
investigate possible primary sources.  
If secondary production dominates, the positron fraction is expected
to fall as a smooth function of increasing energy. 

The energy budget of the Universe can be broken down into baryonic
matter (about 5\%), dark matter 
(about 23\%) and dark energy (about 72\%) (e.g.\cite{kom08}).
Many particle candidates have been proposed for the dark matter
component. The most widely studied are  
the neutralino from supersymmetric models 
(e.g.\cite{jun96}) and the lightest Kaluza Klein particle from extra
dimension models (e.g.\cite{che02}).
The gravitino (e.g.\cite{buc07}) is also an interesting
candidate. 
High energy antiparticles such as
positrons\cite{tyl89,tur90,kam91,kan02,bal02,bal03,ber05,hop07,iba08} and
antiprotons\cite{jun96,bot04}   
can be produced during the annihilation or decay of these dark matter
particles in 
our galaxy.  
In a previous publication\cite{adr08} we presented the
antiproton-to-proton flux ratio in the energy range 1-100 GeV.  
The data follow the trend expected from secondary production
calculations for antiprotons and place significant constraints 
on contributions to the antiproton flux from dark matter particle
annihilations.  
The possible production of positrons from nearby astrophysical
sources, such as pulsars\cite{har87,ato95,chi96,gri04,bue08}, must be
taken into 
account when interpreting potential dark matter signals. 

Cosmic-ray positrons and electrons have been studied mainly by
balloon-borne instruments with correspondingly short 
observation times, and significant atmospheric overburden (for a
review see\cite{mul01}). Their results show large discrepancies,
especially at high 
energies (above 10~GeV). 
This high energy region is the most interesting since the poorly
understood modulation of particle fluxes by the solar wind  
has no relevant effect and possible signatures of primary
components should 
be most evident.  
Although too statistically limited to draw any significant conclusions,
the most recent high energy measurements\cite{bar95,gol96,gas06} 
indicate a flatter positron fraction than expected from secondary
production models.  

\section{Electron and positron identification}

The PAMELA\cite{pic07} apparatus is a system of electronic particle
detectors optimised for the study of antiparticles in the  
cosmic radiation. It was launched from the Bajkonur Cosmodrome on June
15$^{th}$ 2006 on-board  
a satellite that was placed into a 70.0$^\circ$ inclination orbit, at an
altitude varying between 350~km and 610~km. Electrons and positrons
can be reliably distinguished from the other  
cosmic-ray species impinging on PAMELA (mostly protons) by combining
information provided by the different detector components.  
A permanent magnet spectrometer with a silicon tracking system allows
the rigidity  
(momentum / charge, here in units of GV), and sign-of-charge of the
incident particle to 
be determined. The interaction pattern in a imaging silicon-tungsten
calorimeter allows electrons and positrons to be separated from protons. 


The misidentification of electrons and, in particular, protons is
the largest source of background when 
estimating the positron fraction. This can occur if the sign-of-charge
is incorrectly assigned from  
the spectrometer data, or if electron- and proton-like interaction
patterns are confused in the calorimeter data.  
The antiproton-to-electron flux ratio in the cosmic radiation is
approximately 10$^{-2}$ between 1 and 100 GV but can be 
reduced to a negligible level after electrons are selected using
calorimeter information. The proton-to-positron flux  
ratio, however, increases from approximately 10$^{3}$ at 1 GV to approximately 10$^{4}$ at 100 GV. Robust positron 
identification is therefore required, and the residual proton
background must be estimated accurately.  
The imaging calorimeter is 16.3 radiation lengths (0.6 nuclear
interaction lengths) deep, so 
electrons and positrons develop well contained electromagnetic showers
in the energy range of interest.  
In contrast, the majority of the protons will either pass through the
calorimeter as a minimum ionising particle or  
interact deep in the calorimeter. Particle identification based on the
total measured energy and the  
starting point of the reconstructed shower in the calorimeter 
can be tuned to reject 99.9\% of the protons, while selecting $>$ 95\% 
of the electrons or positrons.
The remaining proton contamination in the positron sample can be
eliminated using 
additional topological information, including the lateral and
longitudinal profile of the shower.  
Using particle beam data collected at CERN we have previously
shown\cite{boe06} that less than one proton out of  
100,000 passes the calorimeter electron selection up to 200 GeV/c,
with a corresponding electron selection efficiency  
of 80\%. 

\begin{figure}[ht]
\includegraphics[width=0.9\textwidth]{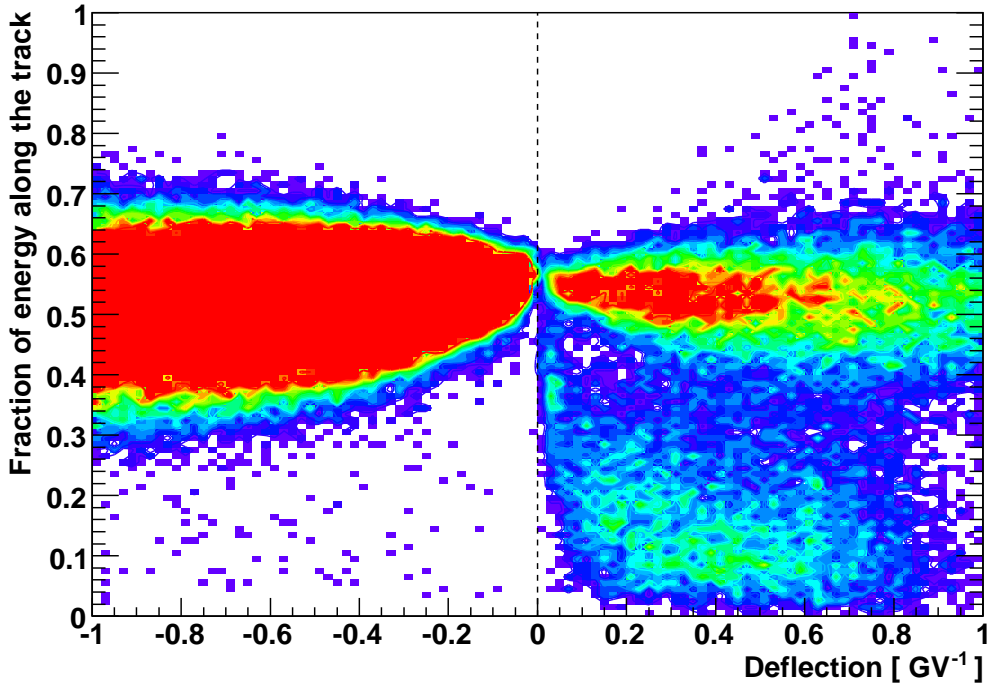}
\caption{{\bf Calorimeter energy fraction  $\cal{F}$.} The 
fraction of calorimeter energy deposited inside a 
cylinder of radius 0.3 Moli\`{e}re radii, as a function of
deflection. The axis of the cylinder is defined by  
extrapolating the particle track reconstructed by the
spectrometer. 
\label{qtrackdefl}
}
\end{figure}

To illustrate this approach, Fig.~\ref{qtrackdefl} shows $\cal{F}$,
the fraction of calorimeter energy deposited inside a  
cylinder of radius 0.3 Moli\`{e}re radii, as a function of deflection
(rigidity$^{-1}$). The axis of the cylinder is defined by  
extrapolating the particle track reconstructed in the
spectrometer. The Moli\`{e}re radius is an important quantity in
calorimetry as it quantifies the 
lateral spread of an electromagnetic shower (about 90\% of the shower
energy is contained in a cylinder with a radius equal to 1 Moli\`{e}re 
radius), and depends only on the absorbing material (tungsten in this
case). The events shown in Fig.~\ref{qtrackdefl}  
were selected requiring a match between the momentum measured by the
tracking system and the total detected energy and the
starting point of the shower in the calorimeter.  
For negatively-signed deflections, electrons are clearly visible as a
horizontal band with $\cal{F}$ lying mostly between 0.4 and 0.7. For
positively-signed deflections,   
the similar horizontal band is naturally associated to positrons, with 
the remaining points, mostly at $\mathcal{F} < 0.4$, designated as proton
contamination.  

The validity of such event characterisations was confirmed using the neutron
yield from the calorimeter 
and the ionization (dE/dx) losses measured in the spectrometer.
These distributions were studied for positively- and
negatively-charged events after the calorimeter selection and  
compared to the corresponding distributions derived from the entire
set of data for negatively charged (mostly electrons) and positively
charged (overwhelmingly proton) events.  
A higher neutron yield is expected in hadronic interactions in the
calorimeter, especially at energies greater than 10~GeV.  
Competing density and logarithmic rise effects for dE/dx losses in the
silicon detectors of the spectrometer 
yield different dE/dx distributions for electrons and protons between 10 and
25~GeV. This is a particularly important check, as the spectrometer
information is independent of the calorimeter  
and can be used to rule out proton interactions 
resulting in (e.g.)
$\pi^{0}$ production in the topmost calorimeter planes. The $\pi^{0}$
will decay to two photons that can generate 
electromagnetic showers in the calorimeter. 
A Kolmogorov-Smirnov test showed that the distributions for 
events characterised as positrons (protons)
were statistically compatible, thereby rejecting the null hypothesis at
5\% level, with the corresponding negatively-
(positively-) 
charged distributions.

The event selection methodology was further validated 
using particle beam data collected prior to launch\cite{boe06} and 
data generated using the PAMELA Collaboration's official simulation
program that reproduces the entire PAMELA apparatus, including the
spectrometer magnetic field and the pressure vessel. Similar
conclusions were derived from cosmic-ray data
collected by the CAPRICE98 balloon-borne 
experiment\cite{boe01}. 
This apparatus was equipped
with a similar but thinner (7 radiation lengths) silicon-tungsten
calorimeter. A gas-RICH detector allowed background-free samples of
protons (i.e. no positron contamination) to be selected up to 50~GeV.
Within the limits of available statistics, the reconstructed proton
and electron/positron lateral energy distributions  
were fully consistent with those obtained with the PAMELA calorimeter. 

\section{Background estimation}

While the distribution shown in Fig.~\ref{qtrackdefl} presents a clear
positron signature, the residual proton  
background distribution must be quantified. 
This distribution was obtained using the flight calorimeter data. There was
no dependence on simulations.  
The total calorimeter depth of 22 detector planes was divided in two
non-mutually exclusive parts:
an upper part comprising planes 1-20,  
and a lower part comprising planes 3-22.
Calorimeter variables (e.g. total detected energy, and lateral shower spread)
were evaluated for both parts. Electrons and positrons can be
identified in the upper part of the calorimeter using the total
detected energy and the 
starting point of the shower. 
The positron component in positively charged events can be
significantly reduced by selecting particles that do not interact in 
the first 2 planes (only 2\% of electrons and positrons with
rigidities greater than 1.5~GV pass this condition).  
This results in a nearly pure sample of protons entering the lower
part of the calorimeter (planes 3-22).  
The procedure was validated using simulations.
As an example Fig.~\ref{presamp}a shows the energy fraction variable,
$\cal{F}$, 
for negatively charged particles in the rigidity range 28--42 GV  
selected as electrons in the upper half of the calorimeter. Panels (b)
and (c) show the $\cal{F}$ distributions for positively-charged
particles obtained for the lower (upper) part of the calorimeter,
i.e. protons (protons and positrons).  
The distributions in panels (a) and (b) are clearly different while 
panel (c) shows a mixture of the two distributions, which  
strongly supports
the positron interpretation for the electron-like $\cal{F}$
distribution in the sample of positively charged events. 
A parametric bootstrap analysis with maximum likelihood fitting was
performed on the distributions shown in Fig.~\ref{presamp} for 
a number of rigidity intervals, and the numbers of detected electrons,
positrons, and contaminating protons were obtained.  
As a cross-check, a non-parametric statistic analysis, using
Kolmogorov-Smirnov tests, were applied to 
electron, positron and proton distributions. The 
results (numbers of identified positrons and protons) were
statistically consistent (well within one standard deviation) with
those obtained using the parametric method. 

\section{The positron fraction}

The results presented here are based on the data-set collected by
PAMELA between July 2006 and February 2008. More than  
10$^9$ triggers were accumulated during a total acquisition time of
approximately 500~days. From these 
triggered events 151,672 electrons and 9,430 positrons were identified
in the energy interval  
1.5 - 100 GeV. Results are presented as a positron fraction
and are shown in
Table~\ref{ratio_table}. The
detection efficiencies for electrons and positrons are assumed to
cancel since the physical processes that these species undergo in the
PAMELA detectors can be assumed to be identical across the energy
range of interest. Possible bias arising from a sign-of-charge
dependence on the acceptance due to the spectrometer magnetic field
configuration and East-West effects caused by the Earth's magnetic
field were excluded as follows.  
Effects due to the spectrometer magnetic field were studied using the
PAMELA Collaboration's simulation  
software. No significant difference was found between the electron and
positron detection efficiency 
above 1 GV. East-West effects as well as contamination from
re-entrant albedo particles (secondary particles 
produced by cosmic-rays interacting with the Earth's atmosphere that
are scattered upward but lack sufficient energy 
to leave the Earth's magnetic field and re-enter the atmosphere in the
opposite hemisphere but at a similar magnetic latitude) are
significant 
around and below the lowest permitted rigidity for a charged
cosmic-ray to reach the Earth 
from infinite distance, known as the geomagnetic cut-off. The
geomagnetic cut-off for the PAMELA orbit varies from less than 100~MV for
the highest   
orbital latitudes to approximately 15~GV for
equatorial regions.  In this work, only events with a measured
rigidity exceeding the 
estimated vertical (PAMELA z-axis) geomagnetic cut-off by a factor of 1.3 were 
considered. This reduced East-West effects and re-entrant particle 
contamination to a negligible amount. The vertical geomagnetic 
cut-off was determined following the St\mbox{\o}rmer formalism
(e.g.\cite{hop64}) on an event-by-event basis and  
using orbital parameters reconstructed at a rate of 1~Hz.

\begin{figure}[ht]
\includegraphics[width=0.9\textwidth]{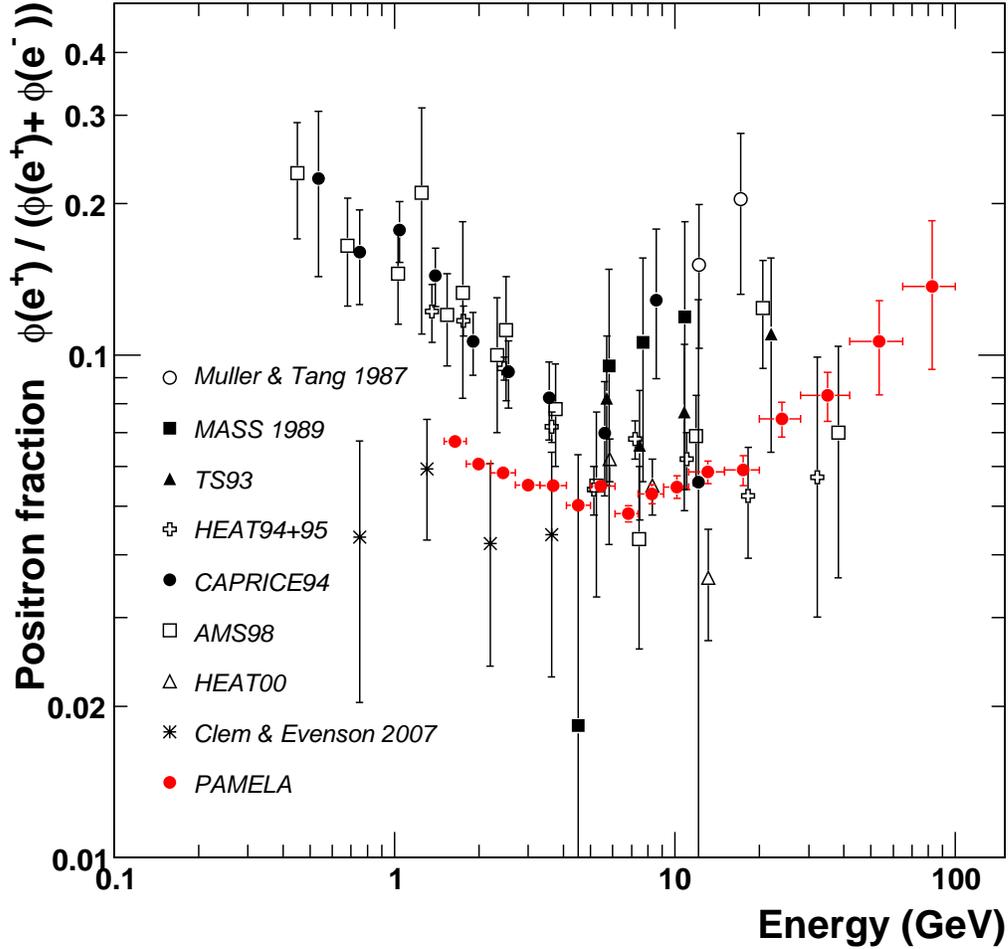}
\caption{{\bf PAMELA positron fraction with other experimental data.}
The positron fraction measured by the PAMELA  
experiment compared with other recent experimental 
data\cite{mul87,gol94,bar97,boe00,alc00,bea04,gas06,cle06}. 
One standard deviation error bars are shown. If not visible, they lie
inside the data points. 
\label{ratio1}}
\end{figure}

Fig.~\ref{ratio1} shows the positron fraction measured by the PAMELA 
experiment compared with other recent experimental data. 
The PAMELA data covers the energy range 1.5 - 100 GeV, with
significantly higher statistics  
than other measurements. 
Two features are clearly visible in the data. At low energies (below
5~GeV) the PAMELA results are systematically 
lower than data collected during the 1990's and at high energies
(above 10~GeV) 
the PAMELA  results show that the positron fraction increases 
significantly with energy.

\section{Observation of charge-sign dependent solar modulation effects}

The solar wind modifies the energy spectra of cosmic-rays within the
solar system. This effect is called solar modulation and can have a
significant effect on cosmic-rays with energies less than about 10~GeV. 
The amount of solar modulation depends on solar activity, 
which has an approximately sinusoidal time dependence and is 
most evident at solar maximum, when the low energy cosmic ray flux is
at a minimum. The peak-to-peak 
period is 11~years, but a complete 'solar cycle' is 22 years long
since at each maximum the polarity of the solar magnetic field 
reverses.  The low energy difference between the PAMELA results and
those from 
CAPRICE94\cite{boe00}, HEAT95\cite{bar97} and AMS-01\cite{alc00}  
are interpreted as a consequence of solar modulation effects. These 
older results were collected during the previous solar cycle  
which favored positively-charged particles due to the solar
polarity. Indications that solar 
modulation effects depend on the cosmic-ray  
sign-of-charge have been clearly seen in the antiproton-to-proton flux
ratio measured before and after the most recent (2000)  
reversal of the solar magnetic field by a series of flights of the
BESS balloon-borne experiment\cite{asa02}. In this case,  
solar modulation effects were seen mostly at low rigidities ($<$2~GV),
and during a period of maximum solar activity. During the period of
solar minimum corresponding to PAMELA data taking, solar modulation
becomes negligible in the BESS data-set. 
The low cosmic-ray antiproton flux limits a detailed study of this effect. 
In the PAMELA data-set, charge dependent solar modulation effects on
electrons and positrons are evident up to nearly 5~GV, even during the
current period of minimum solar activity.  
Contemporary models interpret charge-sign dependent modulation in the
heliosphere as being due to 
gradient, curvature and current sheet drift effects\cite{pot01}. 
Drift effects are at their largest during solar minimum conditions and
mostly affect low mass particles  
such as electrons and positrons, with electrons being favoured in the
current solar cycle. 
A balloon-borne
experiment which flew in June 2006 has also observed a low positron
fraction\cite{cle06} at low energies, but with large statistical
uncertainties.  

\section{The high energy anomaly}

Between 5-10 GeV, the PAMELA positron fraction is compatible with other
measurements.
Previously, the HEAT experiment\cite{cou99} claimed a structure in the
positron 
fraction between 6 and 10~GeV  
but this is not confirmed by the PAMELA data.

\begin{figure}[ht]
\includegraphics[width=0.9\textwidth]{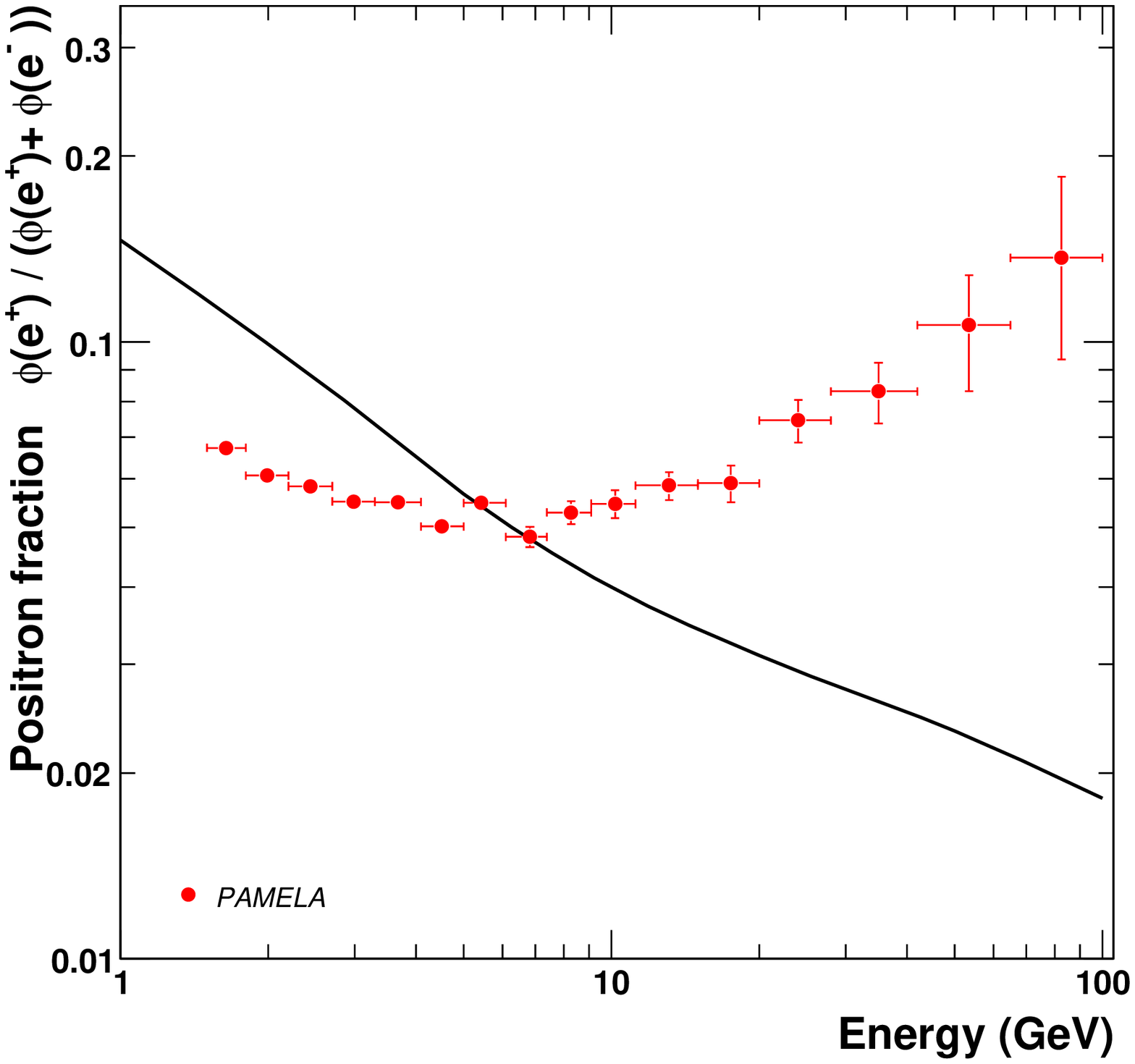}
\caption{{\bf PAMELA positron fraction with theoretical models.}
The PAMELA positron fraction compared with theoretical
model. The solid line shows a calculation by Moskalenko \&
Strong\cite{mos98} for pure secondary production  
of positrons during the propagation of cosmic-rays in the galaxy.
One standard deviation error bars are shown. If not visible, they lie
inside the data points. 
\label{ratio2}}
\end{figure}

Above 10 GeV, the PAMELA results clearly show that the positron fraction
increases significantly 
with energy. Fig.~\ref{ratio2} shows the PAMELA positron fraction
compared to a 
calculation\cite{mos98} for the secondary production of
positrons during the propagation of cosmic-ray nuclei in the galaxy without
reacceleration processes. While this calculation is widely used, it does not
account for uncertainties related to the production of secondary positrons
and electrons. Uncertainties arise due to incomplete knowledge of (a) the
primary cosmic-ray nuclei spectra, (b) modelling of interaction
cross-sections and (c) modelling of cosmic-ray propagation in
the galaxy. Uncertainties on the primary electron spectrum are also
relevant, but since the electron injection spectrum at source is expected to
have a power law index of approximately -2 (e.g.\cite{aha06}) and be equal to
that of protons (e.g.\cite{ber02}) up to about 1 TeV, the positron fraction is
expected to fall as a smooth function of increasing energy if secondary
production dominates.

A rise in the positron fraction at high energy has been postulated for the
annihilation of dark matter particles in the galactic 
halo\cite{tyl89,tur90,kam91,kan02,bal02,bal03,ber05}. The
production of positrons through pair production processes in the
magnetosphere of near-by pulsars would also yield a similar positron
signature\cite{har87,ato95,chi96,gri04,bue08}. 
We note, however, that none of the published models
fit our data well and the reason for the rise remains unexplained.

\section{Conclusions and future prospects}

We have presented the cosmic-ray positron fraction over a wide energy
range, including  
the highest energy ever achieved, and with more than an order of
magnitude increase in statistics over previous experiments. Our
results clearly  
show an increase in the positron abundance  
at high energy that cannot be understood by standard models describing
the secondary production of  
cosmic-rays. 
Either a significant modification in the acceleration and propagation
models for cosmic-rays is needed, or a primary 
component is present. There are several interesting candidates for a
primary component, including the annihilation  
of dark matter particles in the vicinity of our galaxy. There may also
be a contribution from near-by astrophysical sources, such as pulsars.  
The low energy data show a significant charge-sign dependence for solar
modulation and this is the most statistically significant observation
of this effect to date. 
The data are sufficiently precise to allow 
models of the heliosphere to be tuned.  
PAMELA will continue to collect data 
until at least December 2009.
The corresponding increase in statistics will allow higher energies to be
studied (up to the expected spillover limit at approximately 300
GeV). These measurements 
are likely to be important when determining the origin of the observed rise,
especially if an edge is seen in positron fraction as expected in many dark
matter based models. If, on the other hand, the positron fraction is
dominated by a single near-by source, there may be an anisotropy in
the arrival direction of the electrons and positrons\cite{bue08}. 
Work is in progress to reconstruct the positron
fraction down to an energy of 100 MeV, permitting more extensive tests of
solar modulation models.


\begin{figure}[ht]
\includegraphics[width=0.9\textwidth]{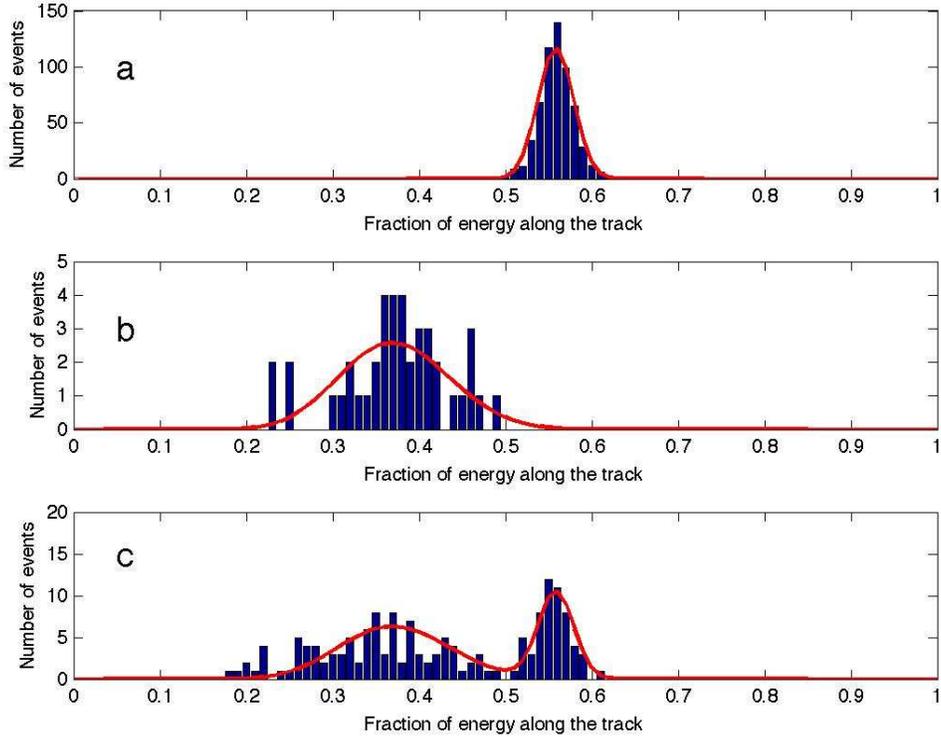}
\caption{{\bf Calorimeter energy fraction  $\cal{F}$: 28--42 GV.} Panel
{\bf a} shows 
the distribution of the energy fraction 
for negatively charged particles, 
selected as electrons in the upper part of the calorimeter. Panel {\bf
b} shows the same 
distribution for 
positively charged particles selected as protons in the bottom
part of the calorimeter. Panel {\bf c} shows positively charged
particles, selected in the 
upper part of the calorimeter, i.e. protons and positrons. \label{presamp}}
\end{figure}

\section{Methods}

\subsection{The PAMELA apparatus.}
\begin{figure}[ht]
\includegraphics[width=0.5\textwidth]{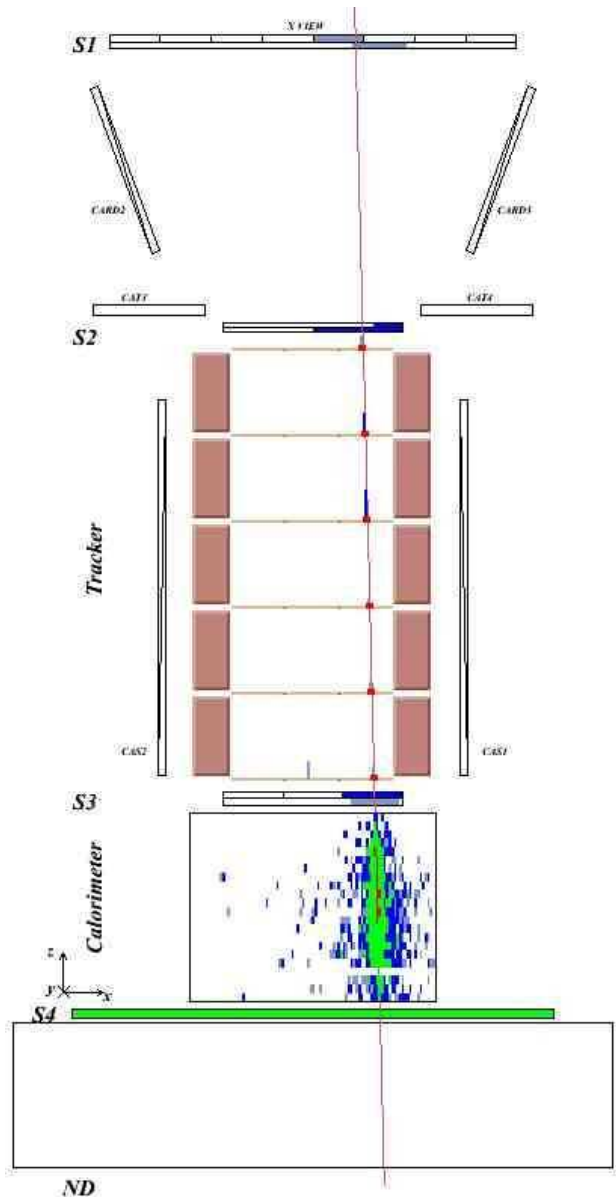}
\caption{{\bf Positron Event display.} A 68 GeV positively-charged
particle selected as positron. The bending (x) view is 
shown. 
The signals as detected by PAMELA
detectors are shown along with the particle trajectory (solid line)
reconstructed by the fitting procedure of the tracking system. The
calorimeter shows the typical signature of an electromagnetic shower
(plane
19 of the calorimeter x-view was malfunctioning).
\label{event}}
\end{figure}

The PAMELA apparatus is inserted inside a pressurized container (2 mm
aluminum window) attached to the  
Russian Resurs-DK1 satellite.
The apparatus, approximately 120~cm tall and with a mass of about
450~kg, can be seen in 
fig.~\ref{event}, which shows a 68 GeV positively-charged particle selected
as a positron. It comprises the following detector systems (from top
to bottom): a time-of-flight system (ToF (S1, S2,
S3)); a magnetic spectrometer; an anticoincidence system (AC (CARD,
CAT, CAS)); an  
electromagnetic imaging calorimeter; a shower tail catcher
scintillator (S4) and a neutron  
detector. The ToF system provides a fast signal for triggering the
data acquisition and  
measures the time-of-flight and ionization energy losses (dE/dx) of
traversing particles. It also allows down-going particles 
to be reliably identified. 
Multiple 
tracks, produced in
interactions above the spectrometer, were rejected by requiring that  
only one strip of the top ToF scintillator (S1 and S2) layers
registered an energy deposition ('hit'). 
Similarly no hits were permitted in either top scintillators of the AC
system (CARD and CAT).  
The central part of the PAMELA apparatus is a magnetic spectrometer
consisting of a 0.43~T permanent magnet  
and a silicon microstrip tracking system. The spectrometer measures
the rigidity 
of charged particles  
through their deflection in the magnetic field. 
During flight the spatial resolution is observed 
to be 3$\mu$m corresponding to a maximum detectable rigidity (MDR)
exceeding 1 TV.  
Due to the finite spatial resolution in the spectrometer, high
rigidity (low deflection) electrons may 'spill over' into the 
positron sample (and vice-versa) if assigned the wrong
sign-of-curvature. This spillover background was eliminated by
imposing a  
set of strict selection criteria on the quality of the fitted
tracks. The spillover limit for positrons is estimated from flight  
data and simulation to be approximately  300~GeV, as expected from 
particle beam tests. 
The dE/dx losses measured in S1 and the 
silicon layers of the magnetic spectrometer were used to select
minimum ionizing singly charged particles (mip)  
by requiring the measured dE/dx to be less than twice that expected from a mip.
The sampling calorimeter comprises 
44 silicon sensor planes
interleaved with 22 plates of tungsten absorber. Each tungsten layer
has a thickness of 0.26~cm corresponding to 0.74 radiation
lengths. Positrons (electrons) can be selected from a background of
protons (antiprotons) by studying the properties of the energy
deposition and  
interaction topology. 
A high dynamic-range scintillator system and a neutron detector are 
mounted under the calorimeter
at the bottom of the apparatus.  

\subsection{Statistical methods.}
For each energy interval, the distribution of the calorimeter energy
fraction ($\mathcal{F}$) for positively-charged
particles (e.g. Fig.~\ref{presamp}c) was expressed as
mixture distribution\cite{eve81} of positrons 
(i.e. signal, electrons as in Fig.~\ref{presamp}a)
and protons (background, e.g. Fig.~\ref{presamp}b):   
\begin{equation}
f(\mathcal{F}) \, = \, p \cdot g_{1}(\mathcal{F};q_{1}) + (1-p) \cdot
g_{2}(\mathcal{F};q_{2}) 
\label{eq}
\end{equation}
where the parameter $p$ gives the mixture proportion;
$g_{1}(\mathcal{F};q_{1})$  and 
$g_{2}(\mathcal{F};q_{2})$ are the probability density functions (p.d.f.) for
positrons/electrons and protons, respectively. 
The p.d.f.'s $g_{1}$ and $g_{2}$ were determined by analysing two
samples of 
pure electrons (Fig.~\ref{presamp}a) and protons (Fig.~\ref{presamp}b)
in the same energy 
range. We used a Beta distribution for both the
electron/positron signal $g_{1}$ and 
for the proton background  
$g_{2}$. In both cases parameter sets $q_{1}$ and $q_{2}$
were determined from a maximum likelihood fit. 

The mixture proportion $p$ was estimated by means of a bootstrap
procedure~\cite{efr79} followed by the maximum likelihood method. 
As first step, the experimental distribution was re-sampled, by means
of a bootstrap procedure,  $N = 1000$ times. 
For each re-sample $i$ ($i = 1,\ldots N$) we estimated the unknown
parameter $p_{i}$ by means of an un-binned maximum likelihood
analysis. The likelihood is defined by: 
\[
L_{i} \, = \, \Pi_{j=1}^{K} [ p_{i} g_{1} 
(\mathcal{F}_{j};q_{1})
+ (1 - p_{i}) g_{2} (\mathcal{F}_{j};q_{2}) ]
\]
where $K$ is the total number of positive particles
(e.g. Fig.~\ref{presamp}c) and j=1,...,K. 

The best fit point for $p_{i}$ corresponds to the maximum $L_{i}$.
Therefore as a result 
we obtained from eq. \ref{eq} 
$N$ estimations of the number of positrons candidates ($n_{i}^{+}$). 
Then, the final number of positron candidates was obtained as 
\[
n \, = \, \frac{1}{N} \sum_{i=1}^{N} n_{i}^{+}
\]
We also estimated the $\alpha$-level confidence interval including all the
values of $n_{i}^{+}$  between the $\alpha$/2 and 1- $\alpha$/2
percentiles of the $n_{i}^{+}$ 
distribution. We chose one standard deviation as the confidence interval. 

An alternative non-parametric statistical method to evaluate the proton
background required the construction of a test sample. The test sample
was 
built by combining the proton sample with a weight $w$ and the electron
sample with a weight $1-w$, with $w\in (0,1)$. The value of $w$ is chosen by
minimizing the Kolmogorov-Smirnov distance between the positive sample and
the test one. Furthermore, a Mann--Whitney test is applied around
the positron peak in order to check if the two sets are compatible. After
the normalization of the test sample, the proton background is found by
counting only the proton events (with their own weight) inside the positron
selection region. The calculation of the confidence interval is based on the
likelihood ratio test\cite{rol05}, by considering proper
probability models for the positron signal, the proton background, the
selection efficiency and the weight $w$.




%


\begin{acknowledgments}
\section{acknowledgments}
 We would like to thank D. Marinucci for helpful discuss
concerning the
statystical methods,  
D. M\"{u}ller and S. Swordy and their
group at University of Chicago, G. Bellettini and G. Chiarelli for
helpful discussion on the data analysis and   
L. Bergstr\"{o}m for useful comments on
the interpretation of our results. 
We acknowledge support from The Italian Space Agency 
(ASI), Deutsches Zentrum f\"{u}r Luft- und Raumfahrt (DLR), The
Swedish National Space  
Board, The Swedish Research Council, The Russian Space Agency
(Roscosmos) and The
Russian Foundation for Basic Research.  

 \textbf{Competing Interests} The authors declare that they have no
competing financial interests.

 \textbf{Correspondence} Correspondence and requests for materials
should be addressed to P.P.~(email: Piergiorgio.Picozza@roma2.infn.it).
\end{acknowledgments}

\begin{table}
\caption{Summary of positron fraction results. The errors are one
standard deviation. \label{ratio_table}} 
\begin{tabular}{ccc}
Rigidity & Mean Kinetic & Extrapolated  \\ 
at & Energy &  \( \frac{\phi(e^{+})}{(\phi(e^{+}) + \phi(e^{-}))} \) \\  
spectrometer & at top of payload  & at top of payload  \\
GV & GeV & \\ \hline
  $1.5$ --  $1.8$ &  $1.64$ & $(0.0673^{+0.0014}_{-0.0013})$ \\
  $1.8$ --  $2.2$ &  $1.99$ & $(0.0607 \pm 0.0012)$ \\
  $2.2$ --  $2.7$ &  $2.44$ & $(0.0583 \pm 0.0011)$ \\
  $2.7$ --  $3.3$ &  $2.99$ & $(0.0551 \pm 0.0012)$ \\
  $3.3$ --  $4.1$ &  $3.68$ & $(0.0550 \pm 0.0012)$ \\
  $4.1$ --  $5.0$ &  $4.52$ & $(0.0502 \pm 0.0014)$ \\
  $5.0$ --  $6.1$ &  $5.43$ & $(0.0548 \pm 0.0016)$ \\
  $6.1$ --  $7.4$ &  $6.83$ & $(0.0483 \pm 0.0018)$ \\
  $7.4$ --  $9.1$ &  $8.28$ & $(0.0529 \pm 0.0023)$ \\
  $9.1$ -- $11.2$ & $10.17$ & $(0.0546^{+0.0029}_{-0.0028})$ \\
 $11.2$ -- $15.0$ & $13.11$ & $(0.0585^{+0.0030}_{-0.0031})$ \\
 $15.0$ -- $20.0$ & $17.52$ & $(0.0590^{+0.0040}_{-0.0041})$ \\
 $20.0$ -- $28.0$ & $24.02$ & $(0.0746 \pm 0.0059)$ \\
 $28.0$ -- $42.0$ & $35.01$ & $(0.0831 \pm 0.0093)$ \\
 $42.0$ -- $65.0$ & $53.52$ & $(0.106^{+0.022}_{-0.023})$ \\
 $65.0$ -- $100.0$ & $82.55$ & $(0.137^{+0.048}_{-0.043})$ \\
\end{tabular}
\end{table}


\end{document}